\documentclass[12pt]{iopart}

\usepackage[cp1251]{inputenc}
\usepackage{iopams}
\usepackage{graphicx,epsfig}
\begin{document}

\title[]{The deterministic dynamics of a single-particle quantum ensemble is equivalent to the stochastic one
due to the indistinguishability of quantum particles}

\author{N. L. Chuprikov}

\address{Tomsk State Pedagogical University, 634041, Tomsk, Russia}
\ead{chnl@tspu.edu.ru} \vspace{10pt}

\begin{abstract}
It is shown that the wave function describing the pure state of a single-particle quantum ensemble, in addition to
statistical restrictions, imposes restrictions on the particle momentum at points in the configuration space
$\mathbb{R}^3$: at time $t$, each point $\mathbf{r}$ is a ``meeting'' point of two (non-interacting) particles of the
ensemble with momenta $\mathbf{p}_1(\mathbf{r},t)$ and $\mathbf{p}_2(\mathbf{r},t)$. Their peculiarity is that the
velocities $\mathbf{p}_1(\mathbf{r},t)/m$ and $\mathbf{p}_2(\mathbf{r},t)/m$ coincide with the velocities
$\mathbf{b}(\mathbf{r},t)$ and $\mathbf{b}_*(\mathbf{r},t)$, which are introduced in Nelson's stochastic approach as key
characteristics of frictionless Brownian particle motion. This means that the instantaneous dynamics of a pair of
non-interacting quantum particles of an ensemble at point $\mathbf{r}$ at time $t$, due to their fundamental
indistinguishability, is equivalent to the collision of two classical Brownian particles. And since this is true at all
times for all points in $\mathbb{R}^3$, the unitary deterministic dynamics of a single-particle quantum ensemble is
equivalent to a stochastic process.
\end{abstract}


\newcommand{\ppp}{\mbox{\hspace{5mm}}}
\newcommand{\ppa}{\mbox{\hspace{15mm}}}
\newcommand{\ppb}{\mbox{\hspace{20mm}}}
\newcommand{\ooo}{\mbox{\hspace{3mm}}}
\newcommand{\ooa}{\mbox{\hspace{1mm}}}

\section{Introduction} \label{int}
A hundred years ago, Schr\"{o}dinger published his famous equation for the wave function $\psi(\mathbf{r},t)$, describing
the state of a spinless, nonrelativistic particle:
\begin{eqnarray}\label{1}
\fl i\hbar\frac{\partial\psi(\mathbf{r},t)}{\partial t}=\hat{H}\psi(\mathbf{r},t).
\end{eqnarray}
The excellent agreement with experiment of all quantum mechanical models subsequently constructed on its basis proved that
this equation adequately describes single-particle quantum dynamics. As a result, it currently plays the same role in
nonrelativistic quantum mechanics as Newton's equation did in classical mechanics. The problem, however, is that the
fundamental equation of motion in classical mechanics is derived from a physical principle (the principle of least
action), while the fundamental equation of motion in quantum mechanics was the fruit of a brilliant guess: Schr\"{o}dinger
did not formulate a physical principle from which this equation could be derived; both the equation itself and the wave
function it obeyed were presented as abstract mathematical constructs lacking any clear physical content.

The first important contribution to this problem was made by Born. Almost immediately after publishing the Schr\"{o}dinger
equation, he proposed a statistical interpretation of the squared modulus of the wave function, as well as a rule for
calculating the expected values of observables for a given wave function, which are valid for the wave function in any
representation. In particular, in the $\mathbf{r}$ representation, $|\psi(\mathbf{r},t)|^2$ is the probability density
function that at time $t$ the particle is in the delta neighborhood of the point $\mathbf{r}$ in the configuration space
$\mathbb{R}^3$ in which the particle moves, and the scalar product
\begin{equation} \label{4a}
\fl \overline{O}_\psi(t)=\int_{\mathbb{R}^3} \psi^*(\mathbf{r},t) \hat{O}\psi(\mathbf{r},t) dx dy dz
\end{equation}
gives the expectation value of an arbitrary observable $O$ at time $t$.

According to Born, quantum mechanics is a statistical theory. To experimentally test its predictions, it is necessary to
conduct (strictly speaking) an infinite number of identical experiments with the same particle. In particular, this is
necessary both to test the formula (\ref{4a}) and to test the predicted probability density $|\psi(\mathbf{r},t)|^2$. That
is, the term ``particle state'' in Born's interpretation should be understood as the state of the corresponding
single-particle quantum statistical ensemble -- an infinite set of identical single-particle systems that are under
identical macroscopic conditions at all moments in time (in the laboratory reference frame). The distribution of particles
in this quantum ensemble over points of the configuration space $\mathbb{R}^3$ is determined by the function
$|\psi(\mathbf{r},t)|^2$.

However, Born left open the question of the physical meaning of the Schr\"{o}dinger equation and the phase of the wave
function. Moreover, a peculiarity of Born's rule (\ref{4a}) is that it coincides with the well-known rule for calculating
the mean value of a random variable in classical probability theory only if the observable $O$ is a function of the
particle's coordinate, that is, if $O=O(\mathbf{r})$. In this case, the operator $\hat{O}$ commutes with the coordinate
operator, and formula (\ref{4a}) takes the form
\begin{equation}\label{4b}
\fl \overline{O}_\psi(t)=\int_{\mathbb{R}^3} O(\mathbf{r})|\psi(\mathbf{r},t)|^2 dx dy dz.
\end{equation}
However, for an observable $O$, the operator of which does not commute with the coordinate operator, the integrand
$\psi^*(\mathbf{r},t) \hat{O}\psi(\mathbf{r},t)$ in (\ref{4a}) remains in the Born interpretation a formal mathematical
expression that has no physical meaning (although the integral (\ref{4a}) itself has a clear physical meaning).

That same year, Madelung published a paper in which he proposed a ``hydrodynamic formulation'' of quantum mechanics. As
was shown, if the wave function is written in the form
\begin{eqnarray}\label{3}
\fl \psi(\mathbf{r},t)=R(\mathbf{r},t)\ooa e^{i\frac{S(\mathbf{r},t)}{\hbar}}\equiv \sqrt{w(\mathbf{r},t)}\ooa
e^{i\frac{S(\mathbf{r},t)}{\hbar}},
\end{eqnarray}
where $S(\mathbf{r},t)/\hbar$ and $w(\mathbf{r},t)$ are real functions, then the Schr\"{o}dinger equation (\ref{1}) can be
written as a system of two real equations that describe the dynamics of an incompressible ``quantum fluid'':
\begin{eqnarray}\label{4}
\fl \frac{\partial w}{\partial t}+\frac{1}{m}\mathbf{\nabla}\left(w\mathbf{\nabla} S\right)=0,
\end{eqnarray}
\begin{eqnarray}\label{5}
\fl \frac{\partial S}{\partial t}+\frac{1}{2m}\left(\mathbf{\nabla}S\right)^2+U_R+V=0;
\end{eqnarray}
where
\begin{eqnarray}\label{6}
\fl U_R=-\frac{\hbar^2}{2mR}\mathbf{\nabla}^2 R\equiv K_w+U_w;\ppp
K_w=\frac{\hbar^2}{8m}\left(\frac{\mathbf{\nabla}w}{w}\right)^2,\ooo U_w= - \frac{\hbar^2}{4m}\frac{\mathbf{\nabla}^2
w}{w}.
\end{eqnarray}

Madelung considered equation (\ref{4}) and the differential consequence of equation (\ref{5}) (which is obtained by
calculating the gradient of the left-hand side of equation (\ref{5})) as a continuity equation and an analogue of Newton's
equation that describe the time variation of the ``quantum fluid'' density $w(\mathbf{r},t)$ ($\int_{-\infty}^\infty
w(\mathbf{r},t) dx dy dz=1$) and its flow velocity $\mathbf{\nabla} S(\mathbf{r},t)/m$ under the action of the external
field $V(\mathbf{r})$ and ``internal'' forces associated with the potential energy $U_R (\mathbf{r},t)$.

Because this interpretation of the Schr\"{o}dinger equation implied a significant departure from the canons of quantum
mechanics, Madelung's hydrodynamic formulation did not gain universal acceptance. Two more versions of the hydrodynamic
formulation were subsequently developed: Louis de Broglie's and Bohm's, of which Bohm's version, known as Bohmian quantum
mechanics, proved more popular.

In this approach, Madelung's ``quantum liquid'' came to be viewed as a model of a single-particle quantum statistical
ensemble. Like Born, Bohm regarded $w(\mathbf{r},t)$ as a function of the distribution of the particles of this ensemble
over the points of the configuration space $\mathbb{R}^3$. However, unlike Born, in addition to the probability density
field $w(\mathbf{r},t)$ in $\mathbb{R}^3$, Bohm introduces a vector field $\mathbf{\nabla} S(\mathbf{r},t)$, which he
views as the field of particle momentum values, and the flow lines of the ``quantum liquid'' as single-particle, nowhere
intersecting trajectories in $\mathbb{R}^3$. Equation (\ref{5}) in Bohm mechanics is an analogue of the Hamilton-Jacobi
equation, which describes the motion of a particle under the influence of an external field $V(\mathbf{r})$ and a
``quantum potential'' $U_R(\mathbf{r},t)$.

However, Bohm's version of the hydrodynamic formulation of quantum mechanics also has not received universal acceptance,
since the interpretation of the value of the vector field $\mathbf{\nabla} S(\mathbf{r},t)$ as a precisely defined value
of the particle's momentum at point $\mathbf{r}$ at time $t$ represents a hidden variable that violates Heisenberg's
uncertainty principle. In this regard, the stochastic formulation of quantum mechanics (see, e.g., \cite{Fen,Nels}), which
respects this principle, is preferable.

In the stochastic approach, the Schr\"{o}dinger equation is derived as an equation describing a random Markov process
representing the frictionless motion of Brownian particles. Unlike Bohmian mechanics, in the Fenyes-Nelson model
\cite{Fen,Nels} the particle trajectories, although continuous, are unpredictable. The process itself is asymmetric with
respect to the flow of time. Therefore, at each point in the configuration space $\mathbb{R}^3$, it is characterized by
two velocities: the forward (in time) velocity $\mathbf{b}(\mathbf{r},t)$ and the backward velocity
$\mathbf{b}_*(\mathbf{r},t)$ (here the notation of \cite{Nels} is used). The average of these two fields gives the field
$\mathbf{\nabla} S(\mathbf{r},t)/m$, and their difference (according to Einstein's theory of Brownian motion) is the
so-called `osmosis velocity':
\begin{eqnarray}\label{1aa}
\fl \frac{1}{2}\left[\mathbf{b}(\mathbf{r},t)+\mathbf{b}_*(\mathbf{r},t)\right]=\frac{\mathbf{\nabla}
S(\mathbf{r},t)}{m},\ppp
\frac{1}{2}\left[\mathbf{b}(\mathbf{r},t)-\mathbf{b}_*(\mathbf{r},t)\right]=\nu\frac{\mathbf{\nabla}
w(\mathbf{r},t)}{w(\mathbf{r},t)},
\end{eqnarray}
where the diffusion coefficient $\nu$ is identified in the work of \cite{Nels} with the quantity $\hbar/2m$. It is
important to note that in this approach, the quantum Bohm ``potential'' is a contribution to the kinetic energy of the
particle.

However, despite the fact that the analogy of single-particle quantum dynamics with the Brownian motion of classical
particles is preferable to the analogy with the motion of an incompressible `quantum fluid', the question of why this
analogy is valid at all remains open. In particular, the Born interpretation (i.e., the one within the framework of
standard quantum mechanics) of the wave function says nothing about the characteristic velocities
$\mathbf{b}(\mathbf{r},t)$ and $\mathbf{b}_*(\mathbf{r},t)$ and what role they might play in the formalism of quantum
mechanics. The very situation to which Nelson's stochastic approach leads is extremely paradoxical: on the one hand, we
are dealing with the Schr\"{o}dinger equation, which describes the deterministic dynamics of a single-particle quantum
ensemble, in which single-particle systems, by definition, do not interact with each other; on the other hand, we are
dealing with equations describing a system of particles colliding with each other in an uncontrolled manner (for the dual
nature of Schr\"{o}dinger dynamics, see \cite{Tak}).

At first glance, these two pictures are incompatible. In any case, if we accept the stochastic interpretation of quantum
mechanics, we must explain the nature of stochasticity in the dynamics of a single-particle quantum ensemble. In
particular, we must explain the role of the velocities $\mathbf{b}(\mathbf{r},t)$ and $\mathbf{b}_*(\mathbf{r},t)$ (key
characteristics of the Brownian motion of particles) in describing the ensemble dynamics within the standard quantum
mechanical formalism.

As we will show below, the solution to these questions lies in the restrictions imposed on the observables' values by the
wave function describing the pure state of the particle. We first considered this issue in the paper \cite{Chu}, where we
proposed a method for determining the fields of observables in the one-dimensional configuration space $\mathbb{R}$ based
on the Schr\"{o}dinger equations and the Born rule for calculating expected observations. Now, generalizing this approach
to the case of the three-dimensional configuration space $\mathbb{R}^3$, we show that the stochastic characteristics
$\mathbf{b}(\mathbf{r},t)$ and $\mathbf{b}_*(\mathbf{r},t)$ arise within the framework of standard quantum mechanics
without any assumptions about the stochasticity of the dynamics of a quantum ensemble of individual particles. Moreover,
using the property of indistinguishability of quantum particles, we answer the question of why the deterministic dynamics
of a single-particle quantum ensemble is equivalent to the Brownian motion of particles.

\section{Fields of observables' operators in the space $\mathbb{R}^3$} \label{observ}

\hspace*{\parindent} Let $O$ be one of the observables of a spinless particle whose operators are defined in the common
for them Schwartz space $S$. If the observable $O$ is a function of the particle coordinate $O(\mathbf{r})$, then its
expectation value is determined by the expression (\ref{4b}), and the function $O(\mathbf{r})$ itself, obviously, is the
field of the operator $\hat{O}$. Thus, the problem is to find fields of observables whose operators do not commute with
the particle coordinate operator $\hat{\mathbf{r}}$. And here it is important to emphasize that the Heisenberg uncertainty
relations impose restrictions on the standard deviations of the position and momentum of a particle in a single-particle
quantum ensemble. In a single-particle experiment, they do not prohibit the particle from having an exact momentum value
at any point of $\mathbb{R}^3$ (see also \cite{Fen}).

For a particle in state $\psi\in S$, given in the coordinate representation, the expected value of such an observable $O$
is given by (\ref{4a}), and the field $O(\mathbf{r},t)$ of the corresponding operator $\hat{O}$ is introduced, according
to \cite{Chu}, by the identity
\begin{equation} \label{4d}
\fl O(\mathbf{r},t)w(\mathbf{r},t)\equiv\mathrm{\Re}[\psi^*(\mathbf{r},t)\hat{O}\psi(\mathbf{r},t)].
\end{equation}

From the mathematical point of view, there are infinitely many other functions that yield the average value of
$\overline{O}_\psi(t)$. These can be obtained by adding to $O(\mathbf{r},t)$ the term
$\mathrm{div}\mathbf{J}(\mathbf{r},t)$, where $\mathbf{J}(\mathbf{r},t)$ is any vector function vanishing at infinity.
But, as shown in \cite{Chu}, if the observable $O$ represents the momentum, kinetic energy, or total energy of the
particle, then it is the definition (\ref{4d}) that yields the field that enters the Schr\"{o}dinger equation and hence
has physical meaning. Moreover, as in the paper by \cite{Chu}, the function $O(\mathbf{r},t)$ for each of the listed
observables will be called for brevity the ``field of the operator $\hat{O}$'' (as shown in the paper by \cite{Chu}, this
is the field of {\it average} values ??of the observable $O$ at points in the configuration space $\mathbb{R}^3$ at time
$t$).

Let's start by defining the momentum operator field. Taking into account Exp. (\ref{3}) for the wave function, we obtain
\begin{eqnarray}\label{8}
\fl
\mathrm{\Re}\left[\psi^*(\mathbf{r},t)\hat{\mathbf{p}}\psi(\mathbf{r},t)\right]=\mathbf{\nabla}S(\mathbf{r},t)w(\mathbf{r},t)\equiv
\mathbf{p}(\mathbf{r},t)w(\mathbf{r},t).
\end{eqnarray}
The vector field $\mathbf{p}(\mathbf{r},t)=\mathbf{\nabla}S(\mathbf{r},t)$ is the desired field of the momentum operator,
since it is precisely this field that appears in equations (\ref{4}) and (\ref{5}). The precise physical meaning of the
field $\mathbf{p}(\mathbf{r},t)$ (and the fields of other operators) will be established in the next section.

Similarly, for the field $E_{kin}$ of the kinetic energy operator we have
\begin{eqnarray}\label{9}
\fl \Re\left[\psi^*(\mathbf{r},t)\frac{\hat{\mathbf{p}}^2}{2m}\psi(\mathbf{r},t)\right]=
\left[\frac{\mathbf{p}^2(\mathbf{r},t)}{2m}+U_R(\mathbf{r},t)\right]w(\mathbf{r},t) \equiv
E_{kin}(\mathbf{r},t)w(\mathbf{r},t),
\end{eqnarray}
and for the Hamiltonian operator field $E(\mathbf{r},t)$ --
\begin{eqnarray}\label{122}
\fl \mathrm{\Re}[\psi^*(\mathbf{r},t)\hat{H}\psi(\mathbf{r},t)]=E_{kin}(\mathbf{r},t)+V(\mathbf{r})\equiv
E(\mathbf{r},t)w(\mathbf{r},t).
\end{eqnarray}

For further consideration, it is advisable to write equations (\ref{4}) and (\ref{5}) as a system of closed equations for
the scalar field $w(\mathbf{r},t)$ and the vector field $\mathbf{p}(\mathbf{r},t)$:
\begin{eqnarray}\label{4n}
\fl \frac{\partial w}{\partial t}+\frac{1}{m}\mathbf{\nabla}\left(w\mathbf{p} \right)=0,
\end{eqnarray}
\begin{eqnarray}\label{5n}
\fl \frac{\partial \mathbf{p}}{\partial
t}+\mathbf{\nabla}\left(\frac{\mathbf{p}^2}{2m}+K_w\right)+\mathbf{\nabla}U_w+\mathbf{\nabla}V.
\end{eqnarray}

\section{On the physical meaning of the fields of momentum and kinetic energy operators, as well as of the ``quantum-mechanical
potential'' $U_R(\mathbf{r},t)$} \label{two}

As we see, from Exp. (\ref{9}) it follows that the function $U_R(\mathbf{r},t)$, like $\mathbf{p}^2(\mathbf{r},t)/2m$, is
part of the field of the kinetic energy operator $E_{kin}(\mathbf{r},t)$. Therefore, Bohm's assumption that the function
$U_R(\mathbf{r},t)=-\frac{\hbar^2}{2mR}\mathbf{\nabla}^2 R=K_w(\mathbf{r},t)+U_w(\mathbf{r},t)$ is a quantum-mechanical
{\it potential}, contradicts the formalism of quantum mechanics. It also follows from here that the fields
$\mathbf{p}(\mathbf{r},t)=\mathbf{\nabla}S(\mathbf{r},t)$ and $E_{kin}(\mathbf{r},t)$ cannot be interpreted as the
momentum and kinetic energy of a particle at the point $\mathbf{r}$ at the time $t$.

Following the paper \cite{Chu}, one might assume that the field $\mathbf{p}(\mathbf{r},t)$ is the average of the two
particle momentum fields $\mathbf{p}_1$ and $\mathbf{p}_2$, and the corresponding two kinetic energy fields give the field
of the kinetic energy operator $E_{kin}(\mathbf{r},t)$:
\begin{eqnarray}\label{4f}
\fl \frac{1}{2}(\mathbf{p}_1+\mathbf{p}_2)=\mathbf{p},\ppp
\frac{1}{2}\left(\frac{\mathbf{p}_1^2}{2m}+\frac{\mathbf{p}_2^2}{2m}\right)= E_{kin}\equiv
\frac{\mathbf{p}^2}{2m}+K_w+U_w.
\end{eqnarray}
However, as shown in \cite{Chu} using the example of a one-dimensional harmonic oscillator in the stationary state (when
${p}(x)\equiv 0$), in this case (due to the contribution of $U_w$) the field $E_{kin}(x)$ takes negative values in
classically forbidden regions of the space $\mathbb{R}$. As a consequence, equations (\ref{4f}) in this case have no real
solutions.

In \cite{Chu}, we interpreted this fact as a requirement of the correspondence principle and did not consider any other
ways of defining the momenta $\mathbf{p}_1$ and $\mathbf{p}_2$. At the same time, such an interpretation of the
correspondence principle is unacceptable for a quantum particle, since the probability of its location in classically
forbidden regions is nonzero. Thus, the momenta $\mathbf{p}_1$ and $\mathbf{p}_2$ must be real in all regions of the
configuration space, and we must take into account the fact that the average value of the kinetic operator in quantum
mechanics is defined in two ways:
\begin{eqnarray*}
\fl \overline{K}_\psi(t)=\frac{1}{2m}\int_{\mathbb{R}^3}\psi^*(\mathbf{r},t)\hat{\mathbf{p}}^2\psi(\mathbf{r},t)
d\mathbf{r} =\frac{1}{2m}\int_{\mathbb{R}^3}|\hat{\mathbf{p}}\psi(\mathbf{r},t)|^2 d\mathbf{r}.
\end{eqnarray*}

Although both of these expressions yield the same value of $\langle K\rangle_\psi$, the second one yields a positive field
$E_{kin}(\mathbf{r},t)-U_w(\mathbf{r},t)=\mathbf{p}^2/2m+K_w$, which differs from the kinetic energy operator field
$E_{kin}(\mathbf{r},t)$ by the function $U_w(\mathbf{r},t)$. And the very fact that the standard formalism of quantum
mechanics allows the second definition of the expectation value $\overline{K}_\psi(t)$ of kinetic energy means that this
function (whose contribution to $\overline{K}_\psi(t)$ is zero) must be ignored when defining the momenta $\mathbf{p}_1$
and $\mathbf{p}_2$ in order to guarantee the existence of a quantum particle's momentum at any point in configuration
space:
\begin{eqnarray}\label{11}
\fl \frac{1}{2}(\mathbf{p}_1+\mathbf{p}_2)=\mathbf{p},\ppp
\frac{1}{2}\left(\frac{\mathbf{p}_1^2}{2m}+\frac{\mathbf{p}_2^2}{2m}\right)= \frac{\mathbf{p}^2}{2m}+K_w.
\end{eqnarray}

In addition, we must take into account Eq. (\ref{5n}) and require that the sought momenta also satisfy the differential
consequence of the second equation in (\ref{11}) --
\begin{eqnarray*}
\fl \frac{1}{2}\mathbf{\nabla}\left(\frac{\mathbf{p}_1^2}{2m}+\frac{\mathbf{p}_2^2}{2m}\right)=
\mathbf{\nabla}\left(\frac{\mathbf{p}^2}{2m}+K_w\right).
\end{eqnarray*}
Given the equality $[\mathbf{v}\times[\mathbf{\nabla}\times \mathbf{v}]]=
\frac{1}{2}\mathbf{\nabla}(\mathbf{v}^2)-(\mathbf{v}\cdot\mathbf{\nabla})\mathbf{v}$ which is true for any vector
$\mathbf{v}$, as well as equalities $[\mathbf{\nabla}\times \mathbf{\nabla} S]=0$ and $[\mathbf{\nabla}\times
\mathbf{\nabla} (\ln w)]=0$, this vector equation is reduced to the form
\begin{eqnarray}\label{13a}
\fl (\mathbf{p}_1\mathbf{\nabla})\mathbf{p}_1+(\mathbf{p}_2\mathbf{\nabla})\mathbf{p}_2 =
2\left[(\mathbf{p}\mathbf{\nabla})\mathbf{p}+(\mathbf{p}_w\mathbf{\nabla})\mathbf{p}_w\right];\ppp
\mathbf{p}_w=\frac{\hbar}{2}\frac{\mathbf{\nabla}w}{w}.
\end{eqnarray}

It is easy to show that the solutions of Eqs. (\ref{11}) and (\ref{13a}) are the vector fields
\begin{eqnarray}\label{14}
\fl \mathbf{p}_1(\mathbf{r},t)=\mathbf{p}(\mathbf{r},t)-\mathbf{p}_w(\mathbf{r},t),\ppp
\mathbf{p}_2(\mathbf{r},t)=\mathbf{p}(\mathbf{r},t)+\mathbf{p}_w(\mathbf{r},t).
\end{eqnarray}
Of importance is that the corresponding velocities
\begin{eqnarray}\label{2aa}
\fl \mathbf{v}_1(\mathbf{r},t)=\frac{\mathbf{p}(\mathbf{r},t)}{m}+\frac{\hbar}{2m}\frac{\mathbf{\nabla}w}{w},\ppp
\mathbf{v}_2(\mathbf{r},t)=\frac{\mathbf{p}(\mathbf{r},t)}{m}-\frac{\hbar}{2m}\frac{\mathbf{\nabla}w}{w}
\end{eqnarray}
coincide with the velocities $\mathbf{b}$ and $\mathbf{b}_*$ introduced in Nelson's stochastic approach \cite{Nels}.

\section{The fields $\mathbf{p}_1(\mathbf{r},t)$ and $\mathbf{p}_2(\mathbf{r},t)$ and the Heisenberg uncertainty relations}
\label{Hei}

Let us show that the fields $\mathbf{p}_1(\mathbf{r},t)$ and $\mathbf{p}_2(\mathbf{r},t)$ satisfy the Heisenberg
uncertainty relations. To do this, we calculate the product of the variances $D_x$ and $D_{p_x}$, which are defined by the
expressions
\begin{eqnarray*}
\fl D_x=\int_{\mathbb{R}^3} \left(x-\langle x\rangle\right)^2 w(\mathbf{r},t)\ooa d\mathbf{r};\ppp \langle
x\rangle=\int_{\mathbb{R}^3} x w(\mathbf{r},t)\ooa d\mathbf{r};
\end{eqnarray*}
\begin{eqnarray*}
\fl D_{p_x}=\int_{\mathbb{R}^3} \left[p_{1,2}^x(\mathbf{r},t)-p^x(\mathbf{r},t)\right]^2 w(\mathbf{r},t)\ooa d\mathbf{r} =
\int_{\mathbb{R}^3} \left[{p}_w^x(\mathbf{r},t)\right]^2 w(\mathbf{r},t)\ooa d\mathbf{r}.
\end{eqnarray*}
Taking into account Exps. (\ref{6}) for the field $\mathbf{p}_w$ and the Cauchy-Bunyakovsky theorem for square-integrable
functions, and integrating by parts, we obtain
\begin{eqnarray}\label{16}
\fl D_{p_x} D_x=\left[\int_{\mathbb{R}^3} \left(\frac{\hbar}{2w}\frac{\partial w}{\partial x}\right)^2 w\ooa
d\mathbf{r}\right] \left[\int_{\mathbb{R}^3} \left(x-<x>\right)^2 w\ooa
d\mathbf{r}\right]\geq\nonumber\\
\fl \frac{\hbar^2}{4}\left[\int_{\mathbb{R}^3} \left(\frac{1}{w}\frac{\partial w}{\partial x}\right)\left(x-<x>\right)
w\ooa d\mathbf{r}\right]^2=\frac{\hbar^2}{4}.
\end{eqnarray}
The inequalities for the variances of the other components of the coordinate $\mathbf{r}$ and the vector fields
$\mathbf{p}_1(\mathbf{r},t)$ and $\mathbf{p}_2(\mathbf{r},t)$ can be obtained similarly.

\section{The role of indistinguishability of quantum particles in the dynamics of a single-particle quantum ensemble}

Thus, we have shown that the wave function describing the pure state of a particle moving in the configuration space
$\mathbb{R}^3$ predicts, in addition to the probability density field $w(\mathbf{r},t)$, two particle momentum fields --
$\mathbf{p}_1(\mathbf{r},t)$ and $\mathbf{p}_2(\mathbf{r},t)$. The peculiarity of these two fields is that the
corresponding velocities $\mathbf{p}_1(\mathbf{r},t)/m$ and $\mathbf{p}_2(\mathbf{r},t)/m$ represent, respectively, the
forward velocity $\mathbf{b}(\mathbf{r},t)$ and the backward velocity $\mathbf{b}_*(\mathbf{r},t)$, characterizing a
random Markov process -- the frictionless motion of Brownian particles. Thus, we have effectively shown that the dynamics
of a single-particle quantum ensemble is equivalent to the dynamics of this Markov process.

But where does stochasticity come from in a single-particle quantum ensemble, whose particles, by definition, do not
interact with each other? The main feature of this question is that in answering it we must reject the assumptions that
stochasticity is caused by vacuum fluctuations etc., since this takes us beyond the framework of standard quantum
mechanics.

To answer this question within the framework of quantum mechanics, we note that in a single-particle quantum ensemble at
any fixed time $t$ in the laboratory reference frame, each point $\mathbf{r}$, according to our approach, is the
intersection point of two locally straight trajectories of particles with momenta $\mathbf{p}_1(\mathbf{r},t)$ and
$\mathbf{p}_2(\mathbf{r},t)$. And since they (by the definition of a single-particle quantum ensemble) do not interact
with each other, these trajectories should also exist at subsequent times. So that, at first glance, one can speak of the
existence of hidden (non-interacting) single-particle trajectories.

However, it is necessary to take into account the fact that we are talking about quantum particles, that is, about
particles that are, in principle, indistinguishable from each other -- the behavior of two indistinguishable particles at
the point of their meeting is equivalent to the behavior of two randomly colliding distinguishable particles participating
in Brownian motion. And since this situation occurs at every point in the configuration space $\mathbb{R}^3$, the dynamics
of a single-particle quantum ensemble as a whole is equivalent to the dynamics of frictionless Brownian motion.

Thus, our approach, developed strictly within the framework of standard quantum mechanics, provides new arguments in favor
of the stochastic interpretation of quantum mechanics. But we now see that the main mystery of quantum mechanics that
needs to be solved lies not in the double-slit experiment, but in the fact that quantum particles are indistinguishable.

\end{document}